\documentclass[a4paper,fceqn]{cas-sc}

\usepackage{amsmath}
\usepackage[numbers]{natbib}

\def\tsc#1{\csdef{#1}{\textsc{\lowercase{#1}}\xspace}}
\tsc{WGM}
\tsc{QE}
\tsc{EP}
\tsc{PMS}
\tsc{BEC}
\tsc{DE}
\begin{document}
\let\WriteBookmarks\relax
\def\floatpagepagefraction{1}
\def\textpagefraction{.001}
\shorttitle{On the structural similarity between liquids and crystals}
\shortauthors{VA Levashov et~al.}

\title [mode = title]{Investigation of the degree of local structural similarity between the parent-liquid and children-crystal states for a model soft matter system}                      
\tnotemark[1]

\tnotetext[1]{This document is the results of the research
   project funded by the Russian Science Foundation (RNF-grant 18-12-00438).}


\author[1,4]{V.A.~Levashov}[style=russian,
                        orcid=0000-0001-7511-2910]
\cormark[1]
\fnmark[1]
\ead{valentin.a.levashov@gmail.com}

\author[2,3,4]{R.E.~Ryltsev}[style=russian,orcid=0000-0003-1746-8200]

\author[4]{N.M.~Chtchelkatchev}[style=russian,orcid=0000-0002-7242-1483]


\address[1]{Technological Design Institute of Scientific Instrument Engineering, 630055, Novosibirsk, Russia.}
\address[2]{Institute of Metallurgy of the Ural Branch of the Russian Academy of Sciences, 620016, Ekaterinburg, Russia}
\address[3]{Ural Federal University, 620002, Ekaterinburg, Russia}
\address[4]{Institute for High Pressure Physics, Russian Academy of Sciences, 108840, Moscow (Troitsk), Russia}

\cortext[cor1]{Corresponding author:}

%

\begin{abstract}
We investigate the degree of local structural similarity between 
the parent-liquid and children-crystal states for a model soft-matter 
system of particles interacting through the harmonic-repulsive pair potential. 
At different pressures, this simple system crystallizes 
into several significantly different crystal structures.
Therefore, the model is well suited for addressing the question under consideration. 
In our studies, we carefully analyze the developments of the pair and triple correlation 
functions for the parent-liquid as the pressure increases. 
In particular, these considerations allow us to address the similarities 
in the orientational orderings of the corresponding liquid and solid phases. 
It is demonstrated that the similarities in the orientational ordering 
between the two states extend beyond the first and second neighbors. 
Currently, it is widely accepted that orientational ordering is important 
for understanding the behaviors of liquids, supercooled liquids, 
and the development of detailed theories of the crystalization process. 
Our results suggest that, up to a certain degree, it might be possible to predict 
the structures of the children-solids from studies of the parent-liquids.
Our results raise anew a general question of how much insight into 
the properties of the liquid-state can be gained from drawing a parallel with the solid-state. 
\end{abstract}

\begin{graphicalabstract}
\includegraphics{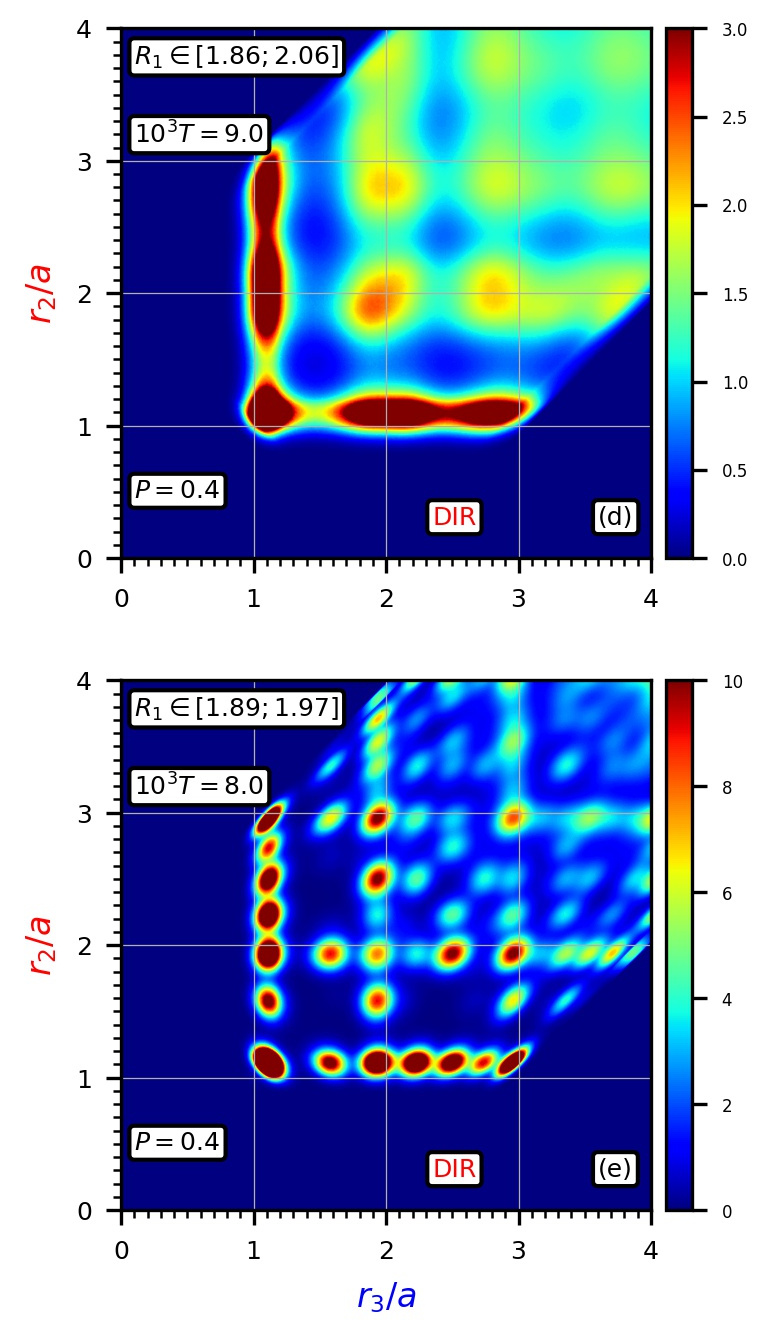}
\end{graphicalabstract}


\begin{highlights}
\item Structural similarity between the parent-liquids and children-crystals states studied
\item Similarity in the orientational ordering extends well beyond the second neighbors
\item Changes in the corresponding liquid and crystal structures are correlated
\item Suggested methods to measure the degree of the structural similarity
\item Suggested applications of the observed structural similarity between the two states
\end{highlights}

\maketitle

\section{Introduction 1 \label{sec:intro1}}

Despite a long history of investigations, there is still no complete understanding 
of the liquids' structures (LSs) at the scale of a few interparticle separations 
and the evolution of these structures with changing external conditions. 
Thus, researchers still develop new and improve existing methods for 
the description of the LSs 
\cite{royall2015role,tong2018revealing,tong2019structural,wei2019assessing,
malins2013identification,zhang2020revealing,lechner2008accurate}.

The difficulties in finding suitable descriptors for the LSs at atomic 
length scales are routinely encountered in the description of LSs of supercooled liquids. 
For example, recent developments of machine learning techniques clearly show 
an existing connection between some structural features 
and dynamics \cite{schoenholz2016structural,boattini2020autonomously,bapst2020unveiling}. 
However, it remains unclear what these structural features are. 
Yet, other approaches show that to understand the connection between 
the structure and dynamics it is necessary 
to consider orientational correlations, i.e., 
to look beyond the pair correlations 
\cite{royall2015role,
tong2018revealing,tong2019structural,
malins2013identification,levashov2020structure}.

The process of crystallization, as multiple recent investigations show,
is also sensitive to the local structural fluctuations that change the degree 
of local orientational ordering 
\cite{kawasaki2010formation,russo2012microscopic,russo2016crystal,
fang2020two,ganapathi2021structure,gebauer2014pre}. 
Thus, it has been demonstrated that crystal nuclei 
form in those regions of the liquid where the orientational ordering 
is more pronounced \cite{kawasaki2010formation,russo2012microscopic,russo2016crystal,ganapathi2021structure}. 
This local orientational ordering in the liquid effectively reduces 
the free energy interface-barrier associated with the nucleus formation \cite{kawasaki2010formation,russo2012microscopic,russo2016crystal,ganapathi2021structure}. 
Further, it has been shown that the structure of the nucleus can 
significantly change as it grows and that the nucleus structure at its 
center can differ from its structure at the interface with the liquid \cite{kawasaki2010formation,russo2012microscopic,russo2016crystal}. 
It is natural to see these results as local structural expressions 
of Ostwald's step rule concerning crystallization 
of polymorphs \cite{Ostwald1897steprule,billinge2009how,ten1999homogeneous,threlfall2003structural}. 
According to this rule, for the polymorph materials, the crystal structure that initially forms from the liquid 
is the structure closest in free energy to the parent liquid, i.e., 
it may not be the most stable (equilibrium) structure at given external conditions.

The mentioned results concerning the nucleus formation and growth processes 
also provoke a general question about the average degree of structural similarity 
between the parent-liquid and children-crystal states.
Thus, while it is known that spatial structural inhomogeneities in liquids 
play an important role in the process of crystallization, it is still of 
interest to clarify how structurally similar are the two phases on average. 
As it follows from the already mentioned publications, this question 
has not been addressed systematically previously. 
However, the actual situation is tangled because of a large amount of related literature.

In our view, the initial vision that structures of parent-liquids 
and children-crystals should be similar, up to a certain degree, can be attributed to the already cited paper by W. Ostwald \cite{Ostwald1897steprule}. 
Then, early structural models of the liquid state often, essentially, were the models of disordered crystals.
A brief account of these approaches is given in the famous J.D. Bernal's paper in which the concept of random close packing (RCP) has been introduced \cite{bernal1959geometrical}. 
The appearance of the RCP model approximately coincided in time with intense developments of computer simulation techniques. Then, researchers mostly simulated atomic systems with strong repulsion at small separations between the particles. For such systems, the RCP model, developed for the hard spheres, leads to a good agreement with the experimental and simulation results. For this reason, the RCP structure essentially became the primary simple model of reference for the liquid state \cite{finney2014renaissance,rowlinson2015rise}. 

We should also note here the following. 
In the times preceding the Bernal's paper~\cite{bernal1959geometrical} 
by at least fifteen years, there had been derived exact relations between the correlation functions describing the structures of liquids.
These equations are known as the Bogoliubov-Born-Green-Kirkwood-Yvon (BBGKY) 
hierarchy~\cite{hansen2013theory,bogoljubov1960problems,bogoliubov1962problems,born1946general,Kirkwood1946,yvon1935theorie}.
Thus, the shift toward consideration of the liquid state without making a connection to the gaseous or solid phases has started even before Ref.~\cite{bernal1959geometrical}.
This shift can also be associated with even earlier works
by L.S. Ornstein and F. Zernike~\cite{rowlinson2015rise,ornstein1914fortuitous,
Ornstein1914Zernike, Ornstein1954Zernike}.
However, it is necessary to remember that BBGKY set of equations 
is incomplete, and it does not allow to gain structural information 
without additional imprecise assumptions, known as closure relations. 
The well-known application of the BBGKY equations allows one 
to calculate the pair density function (PDF) if 
the triple correlation function (TCF), and the interaction potential are known.
However, the TCF is not known usually and thus one has to assume some closure relation between the PDF and TCF. 
The most well-known of these is the Kirkwood's superposition approximation (KSA).
Note also that the knowledge of the PDF calculated 
through the BBGKY or Ornstein-Zernike equations does not provide directly
any knowledge on the orientational ordering in liquids.

The later developments of the DFT methods also lead to numerous investigations of 
the liquids' organizations without discussing 
their possible similarities to some crystal states \cite{hansen2013theory,Likos20011}.

Above, we very briefly described, as we see it, the historical line of the 
investigations of the liquids' structures. This line of progress, in our view, 
led to the situation that there, essentially, were not systematic attempts 
to investigate the degree of the structural similarity between the parent-liquids 
and children-crystal states. One reason that led to this situation is that 
the early studies mostly focused on the liquid systems with strong repulsions, 
i.e., on systems that do not change their structures qualitatively, 
on the increase of pressure, or the decrease in temperature.
However, it is also necessary to point out that there have been more recent 
attempts to draw the parallels between the liquid  
and crystalline states 
\cite{nelson1983order,nelson2002defects,nelson1983liquids,
steinhardt1983bond,
mitus1982theory,mitus1988statistical,mitus1995q446,
patashinski1997towards,son1998modeling,
kivelson1994frustration,
trachenko2008heat,bolmatov2012phonon,trachenko2015collective,
levashov2014understanding,
sun2016crystal}.

In the last three decades, there has been increasing interest in the modeling 
of soft matter systems such as colloids, polymers, micelles, emulsions, and soap bubbles. \cite{evans2019simple,Likos20061,Likos20012,Louis20001,Mohanty20141,mattsson2009soft,
santra2018polymorph,russo2016non,li2016relationship,ouyang2016polymorph,
santra2018polymorph,Levashov20161}. 
The interaction potentials used for the modeling of these systems are significantly 
softer than the potentials used for the modeling of atomic or molecular systems \cite{Likos20012,Louis20001,Mohanty20141}. 
These potentials can even have finite interaction strengths at zero separation between the particles. 
Thus, if these systems are studied at high pressures, a significant overlap between the particles can be observed \cite{Lang20001,Frenkel20091,Prestipino20091,Xu20141,Levashov20161}. 
In this case, there can be non-negligible interactions with the second neighbors; 
this can lead to the formation of substantially different crystal structures, 
at different pressures, even in single-component 
systems of particles with simple interactions \cite{Likos20021,Malescio20071,Levashov20161}. 
For this reason, considerations of soft single-component systems allow investigating 
the degree of the structural similarity between the parent-liquid and children-crystal 
states by comparison of the results at different pressures.

Some authors of this paper previously already discussed the possibility to predict the formation 
of quasicrystalline solids from the shape of the first split peak in the pair density function of the ultrasoft parent-liquid~\cite{ryltsev2015self,ryltsev2017universal}.
Recently, some of us also addressed the structure-structure and structure-property 
relations for a quasicrystal-forming metallic melt system ~\cite{kamaeva2020viscosity}. 

Thus, in this paper, we address the structural similarity between 
the parent-liquids and children-crystals for a soft-matter model system 
consisting of particles interacting through the harmonic-repulsive pair potential. 
This system can be useful for modeling foams, colloids, micelles, and star polymers~\cite{mattsson2009soft,Mohanty20141,Louis20001,Likos20012,Likos20061}.
This and similar systems have been extensively studied previously. 


For the studied system, it has been demonstrated that at different pressures 
it crystallizes into several different crystal structures~\cite{Levashov20161,LuZY20111}.
Therefore, this single component system is very convenient for investigations of the structural 
similarity between parent-liquids and children-crystals. Thus, as we increase the pressure, we
monitor how the liquid's structure changes using the scaled pair density and triple density correlation functions. 
In this process, we compare how the structural changes in the liquid correlate with the changes 
in the previously observed children-crystal states. 
We demonstrate that careful consideration of the pair density function (PDF) 
of the liquid might be sufficient to predict changes in the crystal structures that form from the liquid on cooling.
We also demonstrate that there is a clear similarity between the triple correlation functions (TCFs) 
of the parent-liquid and children-crystal states not only at distances associated with the first and second coordination shells but also at larger distances.

Besides the fact that it is of general interest to understand the structural similarity between parent-liquids 
and children-crystals, this similarity can also be useful for some applications.

For example, to decide which crystal structure is the most stable at given conditions, 
it is necessary to calculate the Gibbs' free energy (GFE) for the chosen crystal lattices. 
But which lattices to consider? For example, the situation with Ref.\cite{Frenkel20091,Prestipino20091,Xu20141,Levashov20161} shows that considerations 
of the narrow sets of the possible lattices can lead to erroneous results. 
On the other hand, to calculate the GFE for every considered lattice, 
using the thermodynamic integration from the reference Einstein crystal 
structure is more demanding than analyzing the structural similarity 
between the liquid and crystal states. 
Thus, this similarity can be useful to decide which lattices 
it is reasonable to consider for further thermodynamic analysis. 

As we show in the paper, careful analysis of the liquid's PDF at different pressures can indicate when a change in the children-crystal structure occurs.

The paper is organized as follows.
In the section~\ref{sec:model} we describe the model and the details of our simulation procedure. 
In section \ref{sec:pdfScaled} we introduce the scaled pair distribution function which is important for our analysis and which precedes  
the related definition of the scaled triple correlation function in \ref{sec:tcfScaled}. 
In sections \ref{sec:datastart}, \ref{sec:tcfScaled}, and \ref{sec:formpartstruct} 
we discuss the obtained data. 
We conclude in section \ref{sec:conclusion}.

\section{The model and details of the simulation procedure \label{sec:model}}

Particles in the studied model interact through the harmonic-repulsive pair potential:
\begin{equation}
  u(r)  =
  \begin{cases}
    \epsilon \left(1-\frac{r}{\sigma}\right)^2, & \text{if $r \leq \sigma$} \\
     0, & \text{if $r > \sigma $} \\
  \end{cases}\label{eq:hrp}
\end{equation}

In our simulations we measure energy in the units of $\epsilon$,
distance in the units of $\sigma$, and time in the units of $\tau = \left(m\sigma^2/\epsilon\right)^{1/2}$.

We used the LAMMPS molecular dynamics package to generate the liquids' structures at different pressures 
and temperatures \cite{Plimpton1995,lammps}. In particular, the particles moved according 
to the Nose-Hoover non-Hamiltonian equations (via the ``npt" and ``iso" commands within the LAMMPS).

Practically, all results reported in this paper have been obtained on the system containing 8000 particles.
Some of the results, obtained on the system of this size, were compared with the data 
collected on the system consisting of 65000 particles.
From these comparisons, which we do not discuss here, we concluded that there are almost 
no size effects in the results presented in this paper.

In simulations, the used value of the time step at $T>0.010$ was $\delta t = 0.001\tau$, while at $T<0.010$ 
the used value of the time-step was $\delta t = 0.010\tau$.
For $T<0.010$ the used value of the Nose-Hoover time-parameter used for the temperature equilibration within the LAMMPS was $1\tau$, 
i.e., 100-time steps, while the used value of the time-parameter for the pressure equilibration was $10\tau$,  i.e., 1000 time steps.
These are the recommended values for these parameters \cite{lammps}.

Initially, we generated the system as the FCC lattice at a very low density of  $\rho=0.04$.
Then, the system was melted and equilibrated at $T=0.015$.
After the equilibration (which happens very fast at $T=0.015$),
the system has been cooled at $P=0.020$ down to $T=0.010$ which is still 
above any observable crystallization temperature for this system.
Then, at $T=0.010$, we increased the pressure from $P=0.020$ to $P=8.0$. For $P<1.0$ the LAMMPS' ``restart" 
files have been saved with the step in pressure $\Delta P = 0.05$, while for $P>1.0$ the ``restart" 
files have been saved with the step in pressure $\Delta P = 0.10$.
Starting from thus obtained restart files, the systems have been equilibrated at all pressures at $T=0.010$.
At this high temperature, the equilibration time at all pressures 
is smaller than $100 \tau$, as can be judged from the dependence of the potential energy on time.

Then, the system(s) at different constant pressures were cooled.
The typical cooling rate used in our simulations was $10^6$ time steps per $\Delta T = 0.001$.
In this paper, we discuss the structures of the liquids at temperatures $T=0.009$, $T=0.007$, and $T=0.006$.
After reaching these temperatures on cooling, systems, at all studied pressures, have been equilibrated for $\sim 10^7$ time steps.
In this equilibration runs and the consequent data collection runs, we monitored the dependence of the mean square displacement on time.
It follows from this monitoring that at all studied pressures and temperatures of the liquids 
the equilibration times are shorter than $10^6$ MD steps, i.e., they are $\approx 10$ times smaller than the times
that we used for the equilibration of the system.
Then, at every temperature and pressure of interest, 100 or 1000 structures have been saved with the time-interval of $10^5$ steps. 
It follows from the analysis of the data that thus produced configurations are sufficient for our purposes here.

The analyses of the generated structures, in all studied cases,
have been made with the self-made programs.

\section{The pair distribution and the scaled pair distribution functions \label{sec:pdfScaled}}

\begin{center}
\begin{table*}
\begin{tabular}{| c | c | c | c | c | c |} \hline
$0.3 < P < 0.7$  & $0.7 < P < 1.7$  & $1.7 < P < 3.9 $  &   $3.9 < P < 6.6$  & $6.6 < P < 6.9$    & $6.9 < P < 8.0$   \\ \hline
$FCC$            & $BCC$            & $Ia\bar{3}d$      & $A5$        &$BCT$            & $P6_{3}/mmc$          \\ \hline
\end{tabular}
\caption{The calculated regions of stability at zero temperature for the lattices  observed in MD simulations. 
The results presented in the table follow from the calculations of the Gibbs free energy at zero temperature~\cite{Levashov20161}. 
The $Ia\bar{3}d$ is a highly symmetric cubic structure with 16 particles in the unit cell. 
The $A5$ is a distorted diamond structure. $BCT$ is a body centered tetragonal lattice. 
The $P6_{3}/mmc$ is the structure formed 
by two alternating triangular lattices. 
See Ref.~\cite{Levashov20161} for more details.}
\label{table:phase-regions}
\end{table*}
\end{center}

A particular idea behind this paper is to consider structural descriptors (SDs) that address 
the changes in the liquid structure that are not reducible to the rescaling of the interparticle distances.
Well-known examples of such SDs are the bond-orientational order parameters (BOOPs) 
and the parameters describing geometries of Voronoi polyhedra.
In this paper, however, we use a simpler approach.
Thus, our analyses partly are based on careful considerations of the pair distribution function (PDF).
The PDF, which is directly related to the scattering experiments, is defined through the pair density $\rho(r)$ as:
\begin{eqnarray}
G(r) \equiv 4\pi r \rho_o \left[g(r) - 1\right],\;\;\;\;\;g(r)\equiv \frac{\rho(r)}{\rho_o},\;\;\;\;
\label{eq:pdfu}
\end{eqnarray} 
where $\rho_o$ is related to the average separation between the particles, $a$: $\rho_o = 1/a^3$. 
Then, $\rho(r)$ is the usual radial pair density.

Further, to address with the PDF those structural changes which 
cannot be reduced to the simple rescaling of the interparticle distances
we also introduce the scaled version of the pair distribution function, $G_{a}(r/a)$:
\begin{eqnarray}
G_{a}(r/a) \equiv 4\pi \left(\frac{r}{a}\right) \left[g_a(r/a) - 1\right],\;\;\;\;g_{a}(r/a) \equiv g(r).\;\;\;
\label{eq:pdfs}
\end{eqnarray} 
It is clear from (\ref{eq:pdfu}) and (\ref{eq:pdfs}) that the relation between the values 
of $G(r)$ and $G_a(r/a)$ is quite simple:
\begin{eqnarray}
G_{a}(r/a) = \left[\frac{1}{\rho_o a}\right] G(r)=a^2 G(r).\;\;\;\;
\label{eq:GGr}
\end{eqnarray} 

\begin{figure*}
\begin{center}
\includegraphics[angle=0,width=6.0in]{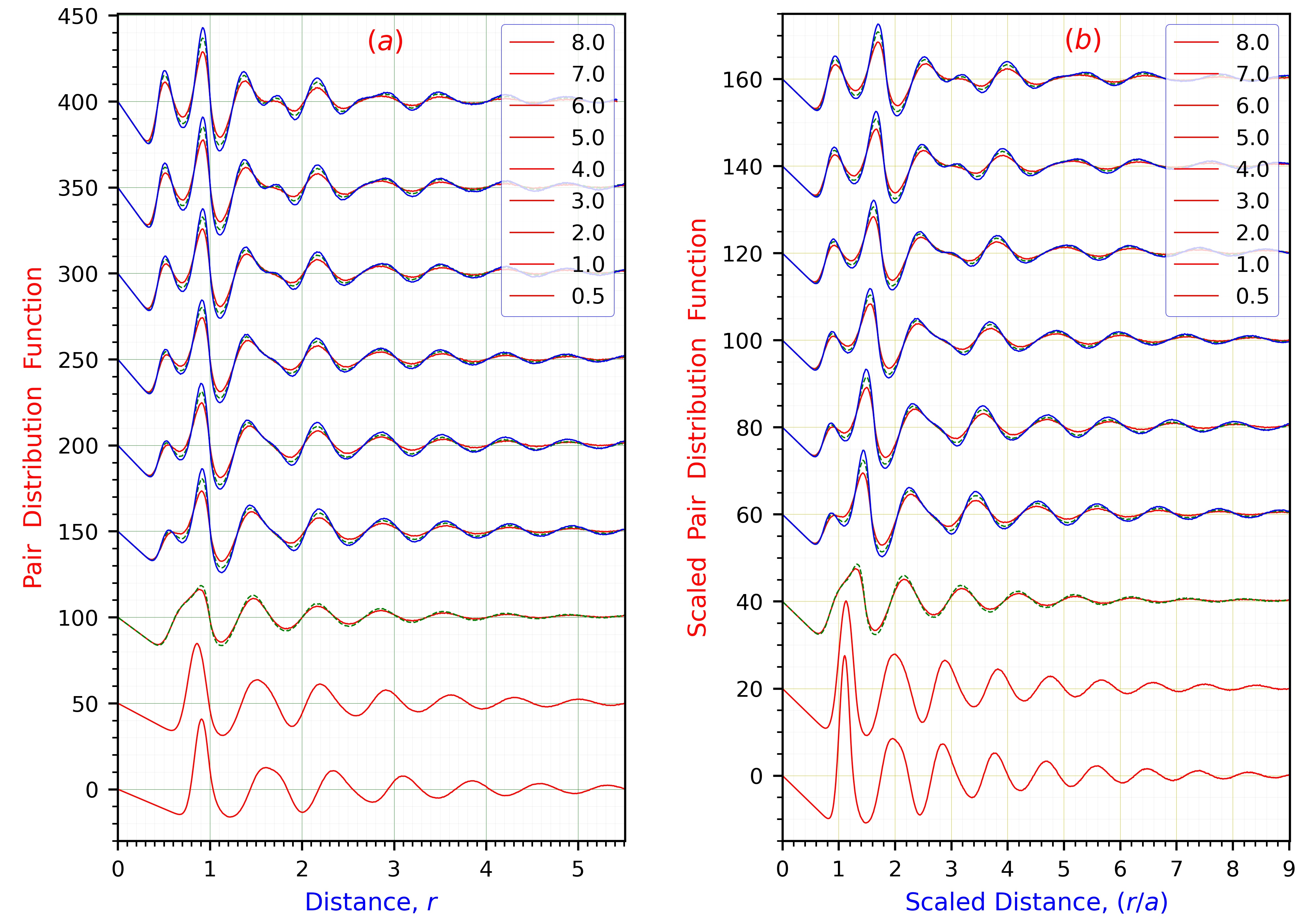}
\caption{
Panel (a) shows the usual, i.e., unscaled, PDFs at the selected pressures and temperatures. 
The red-solid, the green-dashed, and the blue-solid curves present 
the results obtained at $T=0.009$, $T=0.007$, and $T=0.006$
correspondingly.
The curves for the each next higher pressure, $P$, are shifted by 20 upward, 
relative to the results at the each previous lower pressure.
Note that the positions of the second and further peaks, for pressures
$P \geq 2.0$, do not exhibit dependencies on the pressure.
Panel (b) shows the results for the scaled PDFs. 
The curves in panel (b) have been obtained from the same structures 
from which the data in panel (a) have been produced. 
Note that the positions of the second and further 
peaks exhibit clear pressure dependencies at pressures $P \geq 2.0$.
}\label{fig:pdfboth}
\end{center}
\end{figure*}

We already discussed the function $G_{a}(r/a)$ in our 
previous publications \cite{levashov2019anomalous,levashov2020structure}.
Here, we discuss it to demonstrate that behavior of $G_{a}(r/a)$ 
calculated on the liquid state it is sensitive to the change in 
the crystal structures which form from the liquids in the process 
of crystallization.

The structural descriptor (SD) that is central to our further considerations 
is the integral of the square of the $G_{a}(r/a)$:
\begin{eqnarray}
X_a \equiv \int_{\tilde{R}_{min}}^{\tilde{R}_{max}}\left[G_{a}(\tilde{r})\right]^2 d \tilde{r},\;\;\;\;\tilde{r}\equiv r/a.\;\;\;
\label{eq:Xa}
\end{eqnarray} 
In our considerations here, we will assume that the integral extends over all distances for which the pair density function is available.

Note that the integral over the square of the SPDF is related in a simple way to the integral over the square of the non-scaled PDF:
\begin{eqnarray}
X\equiv \int_{R_{min}}^{R_{max}}\left[G(r)\right]^2 d r.\;\;\;
\label{eq:Gr2int}
\end{eqnarray} 

It follows from the previous definitions that:
\begin{eqnarray}
X_a = \left(\frac{1}{\rho_o}\right) X.
\label{eq:Xa-vs-Gr2int}
\end{eqnarray}

\section{Analysis of the obtained data \label{sec:datastart}}

\begin{figure*}
\begin{center}
\includegraphics[angle=0,width=6.0in]{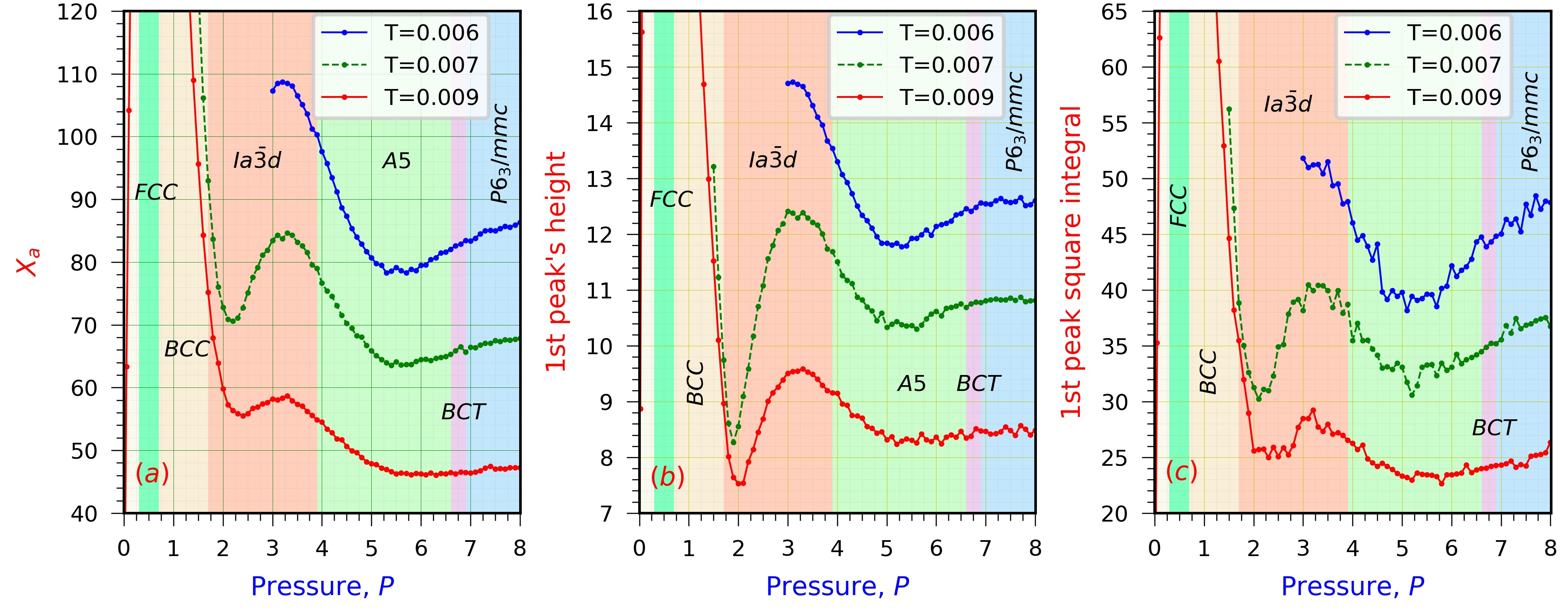}
\caption{
Panel (a) shows the dependencies on the pressure of the integrals (\ref{eq:Xa}). 
The results for the temperatures $T=0.009$, $T=0.007$, and $T=0.006$ are shown for the regions where crystallization were not observed.
The regions of stabilities of the different children-crystalline states are highlighted with different colors.
Panels (b) shows the dependencies on the pressure of the height of the major part of the 1st peak.
Panel (c) shows how the integral over the region of the 1st peak of the square of the scaled PDF depends on the pressure at three studied temperatures.
Note that the curves in all panels exhibit non-monotonous dependencies on the pressure.
}\label{fig:opu}
\end{center}
\end{figure*}

Previously, it has been demonstrated that the system of particles interacting through the 
harmonic-repulsive pair potential crystallizes into several different crystal structures at different pressures. 
These results are summarized in table \ref{table:phase-regions} \cite{Levashov20161}.

In Fig.~\ref{fig:pdfboth} we show the results of calculations 
of the unscaled and scaled PDFs at the selected pressures and temperatures. 
At every pressure, we attempted to consider the liquids at temperatures 
which are sufficiently close to the crystallization transition. 
For the considered ultrasoft system, at small pressures, 
the crystallization transition happens at higher temperatures than at higher pressures. 
While this situation appears to be strange from the perspective of 
the systems with strong diverging repulsions at short separations, 
it is well known for the soft 
systems \cite{Likos20061,Likos20011,LuZY20111,Malescio20071,Frenkel20091,Saija20091}.

At first sight, it appears that the data presented in both panels of Fig. \ref{fig:pdfboth} look quite similar. 
However, there are subtle differences which, in our view, are of importance. 
That is why we suggest that it is reasonable to consider the scaled PDF instead of the unscaled PDF. 
Thus, note in panel (a) that at pressures $P \geq 2.0$ positions 
of the second and further peaks in the unscaled PDFs do not exhibit pressure dependence. 
Thus, it might appear that the density of the system does not change 
in a significant way as pressure increases from $P=2.0$ to $P=8.0$.
However, this impression is completely incorrect. 
Thus, in this interval of the pressure, the density of the system changes 
from $\rho_o \approx 3.2 (1/\sigma^3)$ to $\rho_o \approx 6.5 (1/\sigma^3)$. 
Correspondingly, the average separation between the particles, $a \equiv \rho_o^{-1/3}$, 
changes from $a \approx 0.68 \sigma$ to $a \approx 0.54$. 
This change in density, on the other hand, is easily noticeable 
from the changes in the positions of the large-distance peaks 
in the scaled PDFs shown in panel (b). 
Thus, in our view, the scaled PDF better reflects the structural changes
in the system. For this reason, we use further the scaled PDF 
instead of the unscaled PDF.

Further,  in Panel (a) of Fig.\ref{fig:opu}, we address the behavior of the integral of the square 
of the scaled PDF defined in expressions (\ref{eq:Xa}). 
The first thing to note in the behaviors of $X_a$ is that the shown curves 
exhibit non-monotonous dependencies on the pressure. 
This non-monotonous behavior is more pronounced at lower temperatures than at higher temperatures.
Then, note that qualitatively similar features also could be observed in panels (b,c) that show 
the behaviors of the other, though related, structural descriptors.

Further, we suggest that there is a correlation between the results in Fig.\ref{fig:opu} and in Table \ref{table:phase-regions}.
This observation represents one of the main results of this publication.
Thus, we suggest that the initial rise of the red curve in the region $0.2 \leq P \leq 0.7$ corresponds 
to the formation of the stability region of the FCC lattice. 
Then, the decay of the red curve in the region $0.7 \leq P \leq 1.7$ correlates with the stability region of the BCC lattice. 
In this interval of pressures, according to Fig.\ref{fig:opu}, the amount of the structural order in the system decreases. 
Further, the region $1.7 \leq P \leq 3.9$ is the region of the minimum and the further increase in the structural order.
It is also the region of stability of the $Ia\bar{3}d$ crystal lattice below the crystallization temperature.
At $P \approx 3.9$, the amount of order in the system starts to decrease again in the liquid state. 
At this pressure, a different lattice (distorted diamond) becomes the most stable below the crystallization temperature.
This lattice remains the most stable up to $P \approx 6.6$. 
At $P=6.6$, we do not see a change in the behavior of $X_a$. 
However, $P \approx 6.6$ is already a relatively high pressure. 
At these conditions, the overlap between the particles is quite significant. 
In appears that in such situations, when there occurs a change in the crystal structure, 
the changes in the local orders are not particularly significant. 
This situation concerns not only the neighboring crystalline structures, 
but it is also related to the correlated parent-liquid states from which the crystals form.

Figure \ref{fig:opu} suggests that the transitions from one crystal phase into a different one do not happen near 
the maximums or minimums of the shown ``order parameters" curves. 
Instead, they happen at pressures that are between the maximums and minimums.
The case of the transition from the FCC to the BCC requires more careful considerations.

We summarize the results of the previous two paragraphs in the following way.
In our view, the analysis of the scaled PDFs suggests that the structures
of the studied liquids correlate up to a certain degree with the structures of the crystalline states
which forms from the liquid states in the process of crystallization.

We also note here that similar results could be obtained through the analysis of the unscaled PDF. 
However, we presented the results for the scaled PDF because it better elucidates structural changes 
that are not reducible to the simple rescaling of the interparticle distances. 
The fact that the analyses of the scaled and unscaled PDFs lead to similar results 
shows that the observed features are not artifacts of the behavior of the scaled PDF.

\begin{figure*}
\begin{center}
\includegraphics[angle=0,width=4.8in]{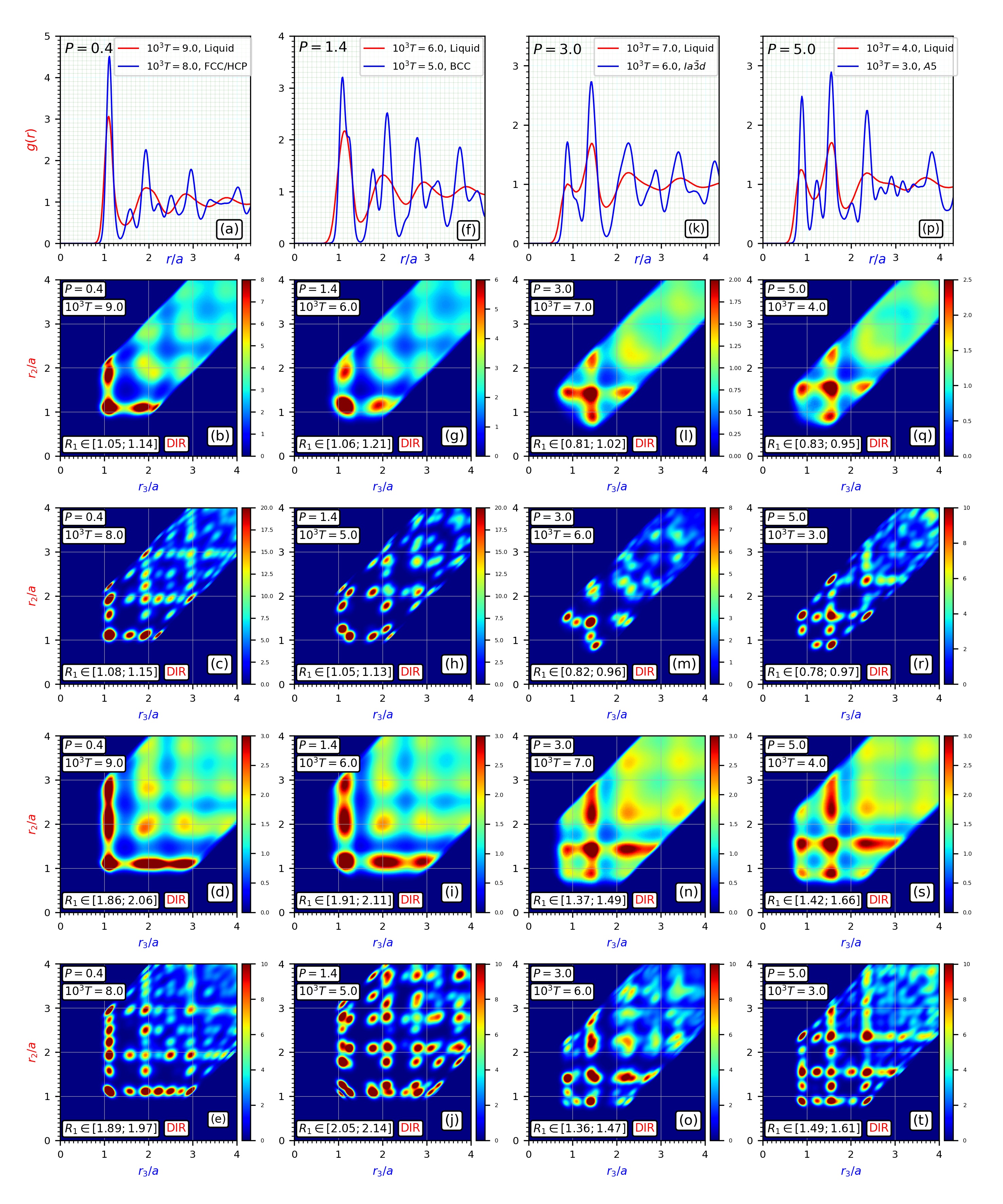}
\caption{
The 1st column of the figure shows the results obtained at $P=0.4$ in the vicinity of the crystallization transition. 
Panel (a) shows the reduced radial pair density functions of the parent-liquid and children-crystalline states. 
Panel (b) shows the slice of the TCF of the liquid that corresponds to the triangles with the length of 
one side approximately equal to the position of the 1st peak in the liquid's PDF in panel (a). 
Panel (c) shows the slice of the TCF of the crystalline state with one argument approximately equal 
to the position of the 1st peak of the PDF of the crystalline state in panel (a). 
Note that there are some similarities in the positions of the intensities in panels (b) and (c). 
Panels (d) and (e) show the slices of the TCFs of the liquid and crystalline states with one argument 
approximately equal to the positions of the second peak in the corresponding PDFs in panel (a). 
The other columns of the figure are similar to the 1st column, but they show the results obtained 
at other selected pressures. See also supplementary materials II for the general explanation of the shape of the TCFs shown in this figure.
}\label{fig:tcf2D1}
\end{center}
\end{figure*}

\section{Analysis with the triple correlation function \label{sec:tcfScaled}}

As we already discussed it previously, the considered system of particles, 
at low enough pressures ($0.1 \leq P < 1.7$), crystallizes into the FCC lattice. 
At higher pressures, 
the system crystallizes into the BCC lattice, 
i.e., in the interval ($1.7 \leq P \leq 2.9$). 
At even larger pressures ($2.0 \leq P \leq 3.9$), 
the liquid crystallizes into the $Ia\bar{3}d$ structure.
In general, it is of interest to investigate if this change in the low temperature crystal structure 
is reflected in the triple correlation function (TCF) of the liquid. 

While the concept of the TCF is well known~\cite{hansen2013theory,Kirkwood1946,born1946general,alder1964triplet,
egelstaff1971experimental,raveche1972three,haymet1981triplet,stillinger1988theoretical,muller1993triplet,ZahnK2003,Ru2003,
coslovich2013static,dhabal2017probing}, 
it is convenient to introduce here its definition in connection 
to the completely random distribution of particles.
Thus, it can be shown that for the completely random distribution of particles in 3D the number of triangles 
with the sides in the ranges $(r_1,\;r_1+dr_1)$, $(r_2,\;r_2+dr_2)$, $(r_3,\;r_3+dr_3)$ {\it per average particle} 
is given by (see Ref.~\cite{levashov2020structure} or Supplementary Materials for this paper):
\begin{eqnarray}
dN_i = 8\pi^2 \rho_o^2 (r_1 dr_1) (r_2 dr_2) (r_3 dr_3).\;\;\;
\label{eq:tcf01}
\end{eqnarray} 
Therefore, for the case of particles characterized by the pair density $\rho(r)$ it is reasonable to assume that:
\begin{eqnarray}
dN_i = \frac{8\pi^2}{\rho_o} \left[\rho(r_1)\rho(r_2)\rho(r_3)\right] (r_1 dr_1) (r_2 dr_2) (r_3 dr_3).\;\;\;
\label{eq:tcf02}
\end{eqnarray} 
Expression (\ref{eq:tcf02}) essentially represents the famous Kirkwood's superposition approximation (KSA) that 
is usually considered as the zero-order reference approximation for the TCF 
of liquids~\cite{hansen2013theory,Kirkwood1946,grouba2004superposition}.
It is well-known that KSA is a rather poor approximation for $dN_i$ in 
real systems \cite{hansen2013theory,alder1964triplet,stillinger1988theoretical,grouba2004superposition}.

It is reasonable to define the TCF in such a way that it is equal to unity (one) for the completely random distribution of particles.
It follows from expression (\ref{eq:tcf01}) that for this we should define the TCF as:
\begin{eqnarray}
C_3(r_1,r_2,r_3) \equiv \frac{dN_i}{(8\pi^2 \rho_o^2)  (r_1 dr_1) (r_2 dr_2) (r_3 dr_3)}.\;\;\;
\label{eq:tcf03}
\end{eqnarray}
Note that (\ref{eq:tcf03}) can also be rewritten in terms of the reduced distances $\tilde{r} \equiv r/a$:
\begin{eqnarray}
\tilde{C}_3(\tilde{r}_1,\tilde{r}_2,\tilde{r}_3) \equiv \frac{dN_i}{(8\pi^2)  (\tilde{r}_1 d\tilde{r}_1) (\tilde{r}_2 d\tilde{r}_2) (\tilde{r}_3 d\tilde{r}_3)}.\;\;\;
\label{eq:tcf04}
\end{eqnarray}
Eventually, it also might be reasonable to consider the function which is equal to zero for the completely random case:
$\tilde{C'}_3(\tilde{r}_1,\tilde{r}_2,\tilde{r}_3) \equiv \tilde{C}_3(\tilde{r}_1,\tilde{r}_2,\tilde{r}_3) - 1$.

We note here that the definition of the TCF that we use is somewhat different from the usually used definition. 
Often, the ratio of the TCF that we defined (\ref{eq:tcf04}) to the TCF which is associated with 
the Kirkwood's superposition approximation (\ref{eq:tcf02}) is regarded as the TCF itself. 
We use definition (\ref{eq:tcf04}) because, for our purposes, it is more convenient to consider 
as the reference state the random distribution of particles instead of considering the non-exiting 
state for which the Kirkwood's approximation is valid.

In our further considerations of the TCF, we will always consider it as a function of the reduced distances. 
Therefore, in the following, we omit the tildes above the arguments of the TCF and above the TCF itself.
We found it convenient to consider the TCF as a function of two arguments $r_2$ and $r_3$ for the (approximately) fixed value of $r_1$.
Few examples of the TCFs for the selected pressures are shown in Fig.~\ref{fig:tcf2D1}.
The results in every column correspond to a particular chosen value of the pressure.

Panels (a,f,k,p) of Fig.~\ref{fig:tcf2D1} show the pair density functions, $g(r)$, of the parent-liquid and children-crystalline
states states in the vicinity of the crystallization transition for the selected pressures.

Panels (b,g,l,q) of Fig.~\ref{fig:tcf2D1} show the ``slices" of the TCFs for the liquid states at the pressures 
that are the same as the pressures in panels (a,f,k,p) correspondingly. 
The ``slices" in panels (b,g,l,q) are the TCFs for the approximately fixed values of the one side, $R_1$, of the triangles. 
These values of $R_1$ correspond to the positions of the 1st peaks of the parent-liquids' $g(r)$ shown in panels (a,f,k,p).

Panels (c,h,m,r) of Fig.~\ref{fig:tcf2D1} show the ``slices" of the TCFs 
for the children-crystalline states at the pressures which are the same 
as the pressures in panels (a,f,k,p) correspondingly. 
The ``slices" in panels (c,h,m,r) are the TCFs for the approximately fixed values of the one side, $R_1$, of the triangles. 
These values of $R_1$ correspond to the positions of the 1st peaks of the children-crystals' $g(r)$ shown in panels (a,f,k,p). 

Note that there is a certain similarity in the positions of the peaks of the $g(r)$ of the liquid and crystalline states in panels (a,f,k,p). 
Then, note that the similarity in the peaks' positions (and in general in the relative color intensities) is also present when we compare
panels (b with c), (g with h), (l with m), and (q with r).
Finally, note how the relative intensities in the liquid and crystalline states change as the pressure changes.

Then, the pairs of panels (d,e), (i,j), (n,o), and (s,t) show the ``slices" of the TCFs of the liquid and 
crystalline states that approximately correspond to the positions of the second peaks in the PDFs in panels (a,f,k,p). 
Note again that there are certain similarities between the denoted pairs of panels.

Thus, the visual analysis of the ``slices" of the TCFs in Fig.~\ref{fig:tcf2D1} shows, beyond doubt, 
that there are structural correlations between the parent-liquid and children-crystalline states. 
Therefore, there appears a natural question: ``how to measure the degree of these correlations?''

Our further quantitative analysis is based on the assumption that the liquid structures, characterized in terms of the reduces TCFs, 
should not change significantly in the interval of pressure at which they crystallize into the same or rather similar crystal structures.
A natural way to measure the similarities in the structures of liquids is through the locations and intensities of the peaks in the TCFs.
Here, under the position of a peak mean its location $(r_1,r_2,r_3)$ in the full TCF (not its ``slices"). 
Thus, if the full TCF is calculated by averaging over all particles in the system and over the different available coordinate snapshots, 
then the local maximums in the TCF can be found straightforwardly. 
For example, if a triangle with the sides $(r_1,\,r_2,\,r_3)$ 
corresponds to the local maximum, then the value of the TCF at 
$(r_1,\,r_2,\,r_3)$ should be the largest in some vicinity of these coordinates, i.e., for example, in the region $(r_1 \pm dr,\,r_2 \pm dr,r_3 \pm dr)$.
We found that the choice of $dr = 0.1$ is reasonably good for this analysis.

\begin{figure*}
\begin{center}
\includegraphics[angle=0,width=6.0in]{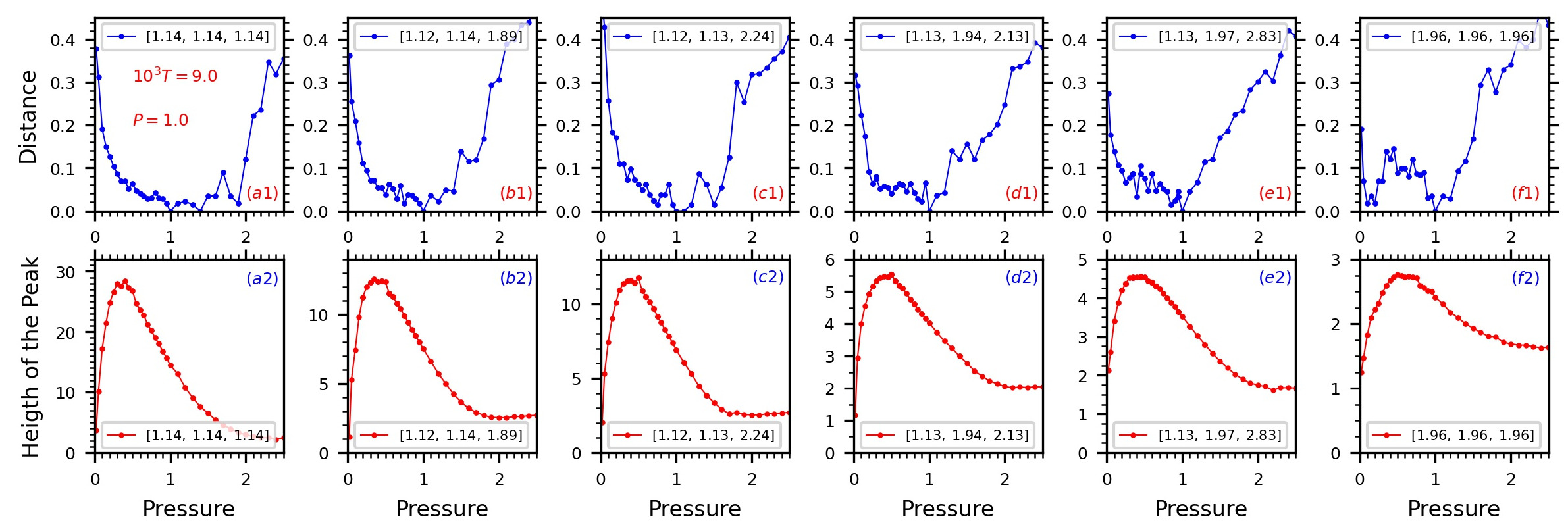}
\caption{At pressure $P=1.0$ and at $10^3 T = 9.0$ the peak with the highest intensity 
in the scaled TCF is located at $\approx\, (1.14,\,1.14,\,1.14)$. 
The ``distance" to this peak from {\it the nearest peak} at other pressures, 
as a function of the pressure at $10^3 T = 9.0$, is shown in panel (a1). 
We see that in the interval of pressures $P \in (0.30,\,1.7)$ the distance 
to the chosen peak located at $\approx\, (1.14,\,1.14,\,1.14)$ is smaller than $0.1$. 
In general, the analysis of the similar other panels and their comparison 
with the results in Table~\ref{table:phase-regions} suggest that 
if the distance to the chosen peak is larger than $\sim 0.15$ then it is 
reasonable to think that the liquid structure has changed sufficiently 
to lead to a different crystal structure in the process of crystallization. 
For example, this happens at $P=1.8$, in agreement with 
the results in Table~\ref{table:phase-regions}.
The case of the transition from the FCC to the BCC structure, 
which happens at $P \sim 0.7$, is an exception that requires more careful analysis.
Panel (a2) of the figure shows how the height of 
the peak nearest to the chosen peak depends on the pressure.
The second highest peak in the TCF at $P=1.0$ and $10^3 T=9.0$ is the peak located at $\approx (1.13,\,1.15,\,1.88)$. 
The behavior of the peaks nearest to it, at other pressures, is shown in panels (b1) and (b2). 
From the behavior of the two parameters associated with this peak, we can conclude again that, 
in the range of pressure $P \in (0.4,\;1.8)$, the structure of the liquid does not change significantly.
The results for the 3rd highest peak located $\approx (1.11,\,1.14,\,2.24)$ also support the suggested conclusion. 
The results for the other three peaks, with noticeably smaller intensities, also support the made conclusion, though somewhat less convincingly. 
The shown curves were obtained by averaging over 1000 frames of 8000 particles.
}\label{fig:peak-P1T9}
\end{center}
\end{figure*}

\begin{figure*}
\begin{center}
\includegraphics[angle=0,width=6.0in]{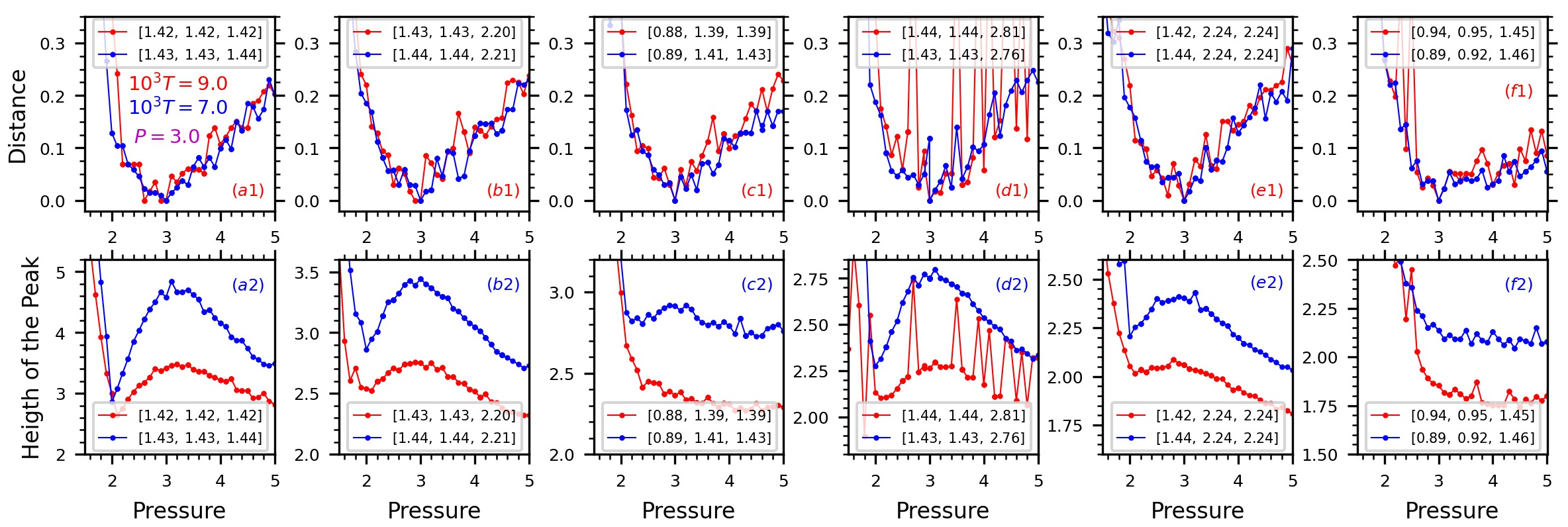}
\caption{Similar to the results in Fig.~\ref{fig:peak-P1T9}, but the chosen peaks were selected at $P=3.0$. 
The results for temperatures $10^3 T = 9.0$ and $10^3 T = 7.0$ are shown.
In the figure, the considered peaks were organized according to their heights at $P=3.0$
In the region of pressures $P \in (1.7,\,3.9)$, according to Table~\ref{table:phase-regions}, 
the liquid crystallizes into the $Ia\bar{3}d$ lattice. 
The data shown in the figure suggest that the region of pressures $P \in (2.0,\,4.0)$ 
indeed can be considered as the region of the relative ``structural stability" for the liquid.
The large fluctuations in the data, in panels (d1) and (d2) for $10^3 T = 9.0$, are due 
to the used algorithm for the determination of the peaks.
Thus, for some pressures, the algorithm does not find the proper 
peak and shows the results for a nearby peak.
}\label{fig:peakpP3T9}
\end{center}
\end{figure*}

Now, let us assume that we would like to compare the structures of the liquid at two different pressures. 
If the structures of the liquid at these pressures are similar, 
besides the simple rescaling of the interparticle distances,
then the positions of the several well-pronounced peaks 
in the scaled TCFs of these two states should be close to each other. 
Also, the intensities of these major peaks should be somewhat similar. 
Thus, to compare the structures of the liquid at these two pressure states through the TCFs, we can proceed as follows. 
For a selected peak in the TCF of one structure, we can find the peak nearest to this selected peak in another structure.
For our present considerations, we define ``the distance" between the two peaks in the TCFs of states $a$ and $b$ 
as: $d \equiv \sqrt{(r_1^b - r_1^a)^2 + (r_2^b - r_2^a)^2 + (r_3^b - r_3^a)^2}$.
Further, we can compare the intensities of the two nearest peaks. 
This analysis can be performed for several selected peaks.

In Fig.~\ref{fig:peak-P1T9},\ref{fig:peakpP3T9} and in their captions, 
we demonstrate the application of the analysis described 
above for the TCFs at two pressures and two temperatures.
Thus in Fig.~\ref{fig:peak-P1T9} we describe the behavior of the selected peaks in the TCF with respect 
to the selected peaks at $P=1.0$ and at $10^3 T = 9.0$. In Fig.~\ref{fig:peakpP3T9} we address the behavior 
of the selected peaks with respect to the reference TCF at $P=3.0$. 
The results for the temperatures $10^3 T = 9.0$ and $10^3 T = 7.0$ are shown.

Concerning the behavior of the curves shows in panels (d1,d2) and (f1,f2) of Fig.~\ref{fig:peakpP3T9}. 
Note that the peaks whose behavior these panels address are relatively week and that in the corresponding 
TCFs there are other peaks in the vicinity of the selected maximums. 
For this reason, the algorithm developed for the search of the maximums may find a different peak in 
the close vicinity of the chosen one. 
This shortcoming of the algorithm causes irregular spikes in panels (d1,d2) and (f1,f2).

\section{Likelihood of formation of a particular structure \label{sec:formpartstruct}}

Another point of interest concerns the possibility to predict which crystal lattice will from the liquid, 
with a known structure, in the process of crystallization. 
Because of a large number of possible crystal lattices, it may not be possible 
to predict which crystal lattice will form precisely. 
However, it might be possible, with a certain degree of confidence, 
to discuss the likelihood of crystallization into a particular crystal lattice.

Thus, if several crystal lattices are considered as the candidates, then it is possible to calculate 
the PDFs and TCFs for these lattices, assuming that particles in the lattices can deviate 
from their equilibrium positions with some probability. 
Then, thus obtained PDFs of the candidate lattices can be directly compared with the PDF of the liquid. 
From these comparisons of the shapes of the PDFs, it might 
be already possible to draw some conclusions, as can be seen from the
comparison of panels (f) and (k) in Fig.~\ref{fig:tcf2D1}.

On the next step, one can compare the TCF of the liquid with the TCFs of the candidate lattices. 
The comparisons of the TCFs are, of course, more complicated than the comparisons of the PDFs.  
In one simple comparison, it is possible to compare the positions of the most intense peaks in the TCFs. 
That is what we do in Tables~\ref{table:FCCtripeaks},\ref{table:BCCtripeaks},\ref{table:IA3Dtripeaks}.

Thus, in Tables~\ref{table:FCCtripeaks},\ref{table:BCCtripeaks},\ref{table:IA3Dtripeaks} 
we show the ``coordinates" and intensities of the well-expressed peaks 
in the TCFs for the liquid and crystalline states that form from the liquids.
We remind here that at pressures $P=0.50$,  $P=1.2$, and  $P=3.0$
the liquid crystallizes into the FCC (FCC/HCP), BCC, and $Ia\bar{3}d$ lattices correspondingly.

The first thing that follows from the comparisons of the data for the corresponding parent-liquids 
and children-crystals states is that for both states there exist quite similar triangles. 
This point becomes particularly clear and unambiguous if one compares 
the data for pressure $P=3.0$ in Table~\ref{table:IA3Dtripeaks} with 
the data for pressures $P=0.5$ and $P=1.5$ in Tables~\ref{table:FCCtripeaks},\ref{table:BCCtripeaks}.
Thus, if one asks the question if the liquid at $P=3.0$ is more likely 
to crystallize into the FCC, BCC, or $Ia\bar{3}d$ crystal structure, then the answer is obvious. 
Note that the way we look at the data now is closely related 
to our discussions in the context of Fig.~\ref{fig:tcf2D1}.

Another point concerns subtle differences in the liquids' structures at $P=0.5$ and $P=1.5$.  
First, notice that at $P=0.5$ the peak in the TCF that corresponds to the smallest triangles 
in the system is located at $(1.10,\,1.11,\,1.11)$, i.e., it corresponds to the nearly equilateral 
triangle, as one would expect to be reasonable for the formation of the FCC lattice.
On the other hand, at $P=1.5$, the peak that corresponds to the smallest triangles in 
the liquid state is located at $(1.13,\,1.13,\,1.18)$. 
While the deviation from the completely equilateral triangle for this peak is small, it, nevertheless, is present, 
and one might consider it as an indication against the formation of the FCC lattice.

Another subtle difference can also be noticed with respect to the 4th
lines in Tables~\ref{table:FCCtripeaks},\ref{table:BCCtripeaks}. 
Thus, it follows from these lines that the data for the liquid's state in Table~\ref{table:FCCtripeaks} 
correspond (arguably) to the data for the crystal state in Table~\ref{table:BCCtripeaks} 
to a lesser degree than the data for the liquid's state in Table~\ref{table:BCCtripeaks} itself. 
While these differences are small we still consider them 
as weak footprints of the underlying crystalline states 
imposed on the structures of the liquids' states.

In general, as follows from Fig.~\ref{fig:tcf2D1}, 
the differences between the TCFs for the liquids' and crystalline states, at pressures 
$P=0.5$ and $P=1.4$, are quite subtle and establishment of 
the criteria for the formation of the FCC or BCC lattice from the liquid 
might require quite a different type of consideration.

\begin{center}
\begin{table}
\begin{tabular}{| c | c |} \hline
                                      &                                     \\
Liquid at $P=0.50$                    & FCC crystal                         \\ 
                                      &                                     \\
 $(1.10,\;1.11,\;1.11),\;\;[27.1]$    &  $(1.11,\;1.12,\;1.12),\;\;[60.9]$  \\
 $(1.11,\;1.11,\;1.87),\;\;[13.2]$    &  $(1.11,\;1.12,\;1.95),\;\;[46.5]$  \\                                                                          
 $(1.09,\;1.10,\;2.18),\;\;[11.4]$    &  $(1.11,\;1.13,\;2.24),\;\;[78.9]$  \\        
 $(1.10,\;1.90,\;2.17),\;\;[5.60]$    &  $(1.12,\;1.94,\;2.24),\;\;[18.4]$  \\  
 $(1.11,\;1.96,\;2.85),\;\;[4.74]$    &  $(1.11,\;1.93,\;2.50),\;\;[18.0]$  \\    
 $(1.10,\;2.08,\;2.78),\;\;[4.70]$    &  $(1.10,\;1.94,\;2.73),\;\;[10.0]$  \\  
 $(1.11,\;2.81,\;2.84),\;\;[4.48]$    &  $(1.10,\;2.75,\;2.96),\;\;[13.2]$  \\                           
 $(1.91,\;1.92,\;1.92),\;\;[2.94]$    &  $(1.93,\;1.93,\;1.93),\;\;[11.5]$  \\                           
                                      &                                     \\  \hline  
\end{tabular}
\caption{The 1st column shows the highest peaks in the TCF of the liquid, at $P=0.50$ and $10^3 T = 9.0$, 
in the order of decrease of their magnitude. 
At $P=0.50$ the liquid crystallizes into the FCC lattice.
The numbers in the round and square brackets show the ``coordinates" of the peaks and their heights. 
The 2nd column shows the selected peaks in the TCF of the FCC lattice with the Gaussian particles' displacements of $\sigma = r_{nn}/20$. 
These peaks were selected to correspond approximately to the peaks in the 1st column. 
Note that the second column does not show all peaks that are present in the TCF 
of the FCC lattice and that the shown peaks are not ordered, as the peaks in the 1st column.}
\label{table:FCCtripeaks}
\end{table}
\end{center}


\begin{center}
\begin{table}
\begin{tabular}{| c | c |}                                                     \hline
                                      &                                     \\
        Liquid at $P=1.50$            &              BCC crystal            \\ 
                                      &                                     \\
 $(1.13,\;1.13,\;1.18),\;\;[7.13]$    &  $(1.08,\;1.08,\;1.25),\;\;[31.0]$  \\
 $(1.17,\;1.17,\;1.99),\;\;[4.13]$    &  $(1.09,\;1.09,\;2.18),\;\;[62.3]$  \\                                                                       
 $(1.12,\;1.14,\;2.25),\;\;[3.67]$    &  $(1.25,\;1.25,\;2.51),\;\;[32.2]$  \\        
 $(1.15,\;2.00,\;2.00),\;\;[3.05]$    &  $(1.25,\;2.09,\;2.09),\;\;[18.1]$  \\  
 $(1.14,\;2.08,\;2.98),\;\;[2.56]$    &  $(1.08,\;2.08,\;2.80),\;\;[19.7]$  \\ 
 $(1.13,\;2.02,\;2.86),\;\;[2.48]$    &  $(1.24,\;2.10,\;2.74),\;\;[15.7]$  \\  
 $(1.08,\;2.02,\;2.59),\;\;[2.14]$    &  $(1.09,\;1.77,\;2.74),\;\;[18.1]$  \\                           
 $(2.07,\;2.07,\;2.09),\;\;[2.07]$    &  $(2.10,\;2.12,\;2.12),\;\;[15.4]$  \\                           
                                      &                                     \\  \hline  
\end{tabular}
\caption{The 1st column shows the highest peaks in the TCF of the liquid at $P=1.50$ and $10^3 T = 9.0$ 
in the order of decrease of their magnitude. 
At $P=1.50$ the liquid crystallizes into the BCC lattice.
In other respects, this table is similar to table~\ref{table:FCCtripeaks}.}
\label{table:BCCtripeaks}
\end{table}
\end{center}


\begin{center}
\begin{table}
\begin{tabular}{| c | c |}                                                     \hline
                                      &                                     \\
        Liquid at $P=3.0$             &         $Ia\bar{3}d$ crystal        \\ 
                                      &                                     \\
 $(1.43,\;1.43,\;1.44),\;\;[4.58]$    &  $(1.41,\;1.41,\;1.54),\;\;[24.2]$  \\
 $(1.44,\;1.44,\;2.21),\;\;[3.44]$    &  $(1.40,\;1.40,\;2.35),\;\;[21.2]$  \\                                                             
 $(0.89,\;1.41,\;1.43),\;\;[2.92]$    &  $(0.89,\;1.40,\;1.40),\;\;[13.6]$  \\     
 $(1.46,\;1.46,\;2.87),\;\;[2.75]$    &  $(1.40,\;1.53,\;2.87),\;\;[18.1]$  \\  
 $(1.43,\;1.43,\;2.76),\;\;[2.74]$    &  $(1.54,\;1.54,\;2.52),\;\;[19.7]$  \\     
 $(1.44,\;2.24,\;2.24),\;\;[2.40]$    &  $(1.54,\;2.35,\;2.36),\;\;[15.7]$  \\  
 $(0.89,\;0.92,\;1.46),\;\;[2.13]$    &  $(0.63,\;0.63,\;1.26),\;\;[17.8]$  \\                           
 $(0.87,\;1.42,\;2.28),\;\;[2.00]$    &  $(0.71,\;1.40,\;2.11),\;\;[11.4]$  \\    
 $(2.26,\;2.26,\;2.26),\;\;[1.70]$    &  $(2.33,\;2.33,\;2.33),\;\;[7.21]$  \\                             
                                      &                                     \\  \hline  
\end{tabular}
\caption{The 1st column shows the highest peaks in the TCF of the liquid at $P=3.0$ and $10^3 T = 7.0$ 
in the order of decrease of their magnitudes. 
At $P=3.0$, the liquid crystallizes into the $Ia\bar{3}d$ crystal lattice.
In other respects this table is similar to tables~\ref{table:FCCtripeaks},\ref{table:BCCtripeaks}.}
\label{table:IA3Dtripeaks}
\end{table}
\end{center}


\section{Conclusion \label{sec:conclusion}}

In this paper, we addressed the issue of the local structural similarity 
between the parent-liquid and children-crystal states for a model single component system of particles interacting through the harmonic-repulsive pair potential. 
Our choice of the system originates in the previous works 
that demonstrated that this single component system, at different pressures, 
forms several quite different crystal structures.
This system is suitable for modeling foams, colloids, micelles, and star polymers.

In our investigations, we studied the presence of the structural similarity using two well-known 
tools often used in addressing the structures of liquids and crystals. 
In particular, we considered the pair distribution function (PDF) and the triple correlation function (TCF). 

It follows from the obtained results that there are well-expressed local structural similarities 
between the paren-liquid and the children-crystalline states and that these 
similarities can be revealed using the PDF and the TCF. 
For the observation of these similarities, it is only necessary to perform 
a relatively careful analysis of these correlation functions.

Since the PDF can be accurately measured in experiments, the obtained results suggest 
that some of the developed ideas can be tested experimentally.

In our view, the observations presented in this paper can be useful for the development of new materials. 
For example, let us suppose that the composition of some alloy melt is varied in steps through 
an addition of a particular component. 
Then, our results suggest that from the measurements of the PDF of the melt, it might 
be possible to predict the regions of composition at which the structures 
of the corresponding crystalline states remain stable or change. 

The observed similarities between the parent-liquid and children-crystal states 
also can be used for deciding which crystal lattices it is reasonable to consider 
as viable structures for further considerations using free energy calculations. 
For example,  like those performed in Ref.~\cite{Frenkel20091,Xu20141}.

Our results also suggest that it might be possible to predict, up to a certain degree, 
the properties of the crystalline states from the properties of their melts. 
The model that we studied here is, however, quite specific because it is an ultrasoft system.
Thus, further studies of other systems are needed to understand the possible extent of 
the correlations between the liquid and crystalline structures 
and feasible applications of the observations.

\section{CRediT author statement}

Valentin A. Levashov: Conceptualization, Methodology, Software, Validation, Formal analysis, Investigation, Visualization, Writing - Original Draft, Writing - Review \& Editing. Roman E. Ryltsev and Nikolay M. Chtchelkatchev:  Conceptualization, Methodology, Resources, Project administration, Funding acquisition, Writing - Review \& Editing.

\section{Acknowledgements} 

This work was supported by the Russian Science Foundation (grant 18-12-00438). 
We gratefully acknowledge access to the following computational resources: 
Supercomputing Center of Novosibirsk State University (http://nusc.nsu.ru), 
the federal collective usage center ‘Complex for Simulation and Data Processing 
for Mega-science Facilities’ at NRC ‘Kurchatov Institute’ (http://ckp.nrcki.ru/), 
supercomputers at Joint Supercomputer Center of Russian Academy 
of Sciences (http://www.jscc.ru), and ‘Uran’ supercomputer 
of IMM UB RAS (http://parallel.uran.ru).

\section{Supplementary Materials. \label{sec:supplementary}}

\subsection{Supplementary Materials I. The triple correlation functions.
 \label{sec:suppl-1-triple}\\}

In this Supplementary Materials, for illustrative purposes, 
we describe the general shape of the triple correlation function 
for the two simple archetypical systems, i.e., for the random system 
of particles and the random system of particles with an exclusion distance.\\

We find it convenient to introduce the triple correlation function (TCF) as follows.

\begin{figure}
\begin{center}
\includegraphics[angle=0,width=2.0in]{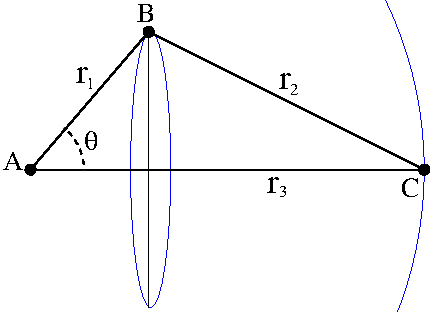}
\caption{
An image for the definition of the triple correlation function (TCF).
See Eq.~\ref{eq:triple1}.
}\label{fig:trianglefig}
\end{center}
\end{figure}

Let us first consider the case of randomly distributed particles.
Let particle ``$A$" be one of the vertexes of the triangle ``ABC", 
as shown in Fig. \ref{fig:trianglefig}.
It is easy to see that the number of triangles, with respect 
to a given particle ``A", with the sides of lengths $(r_1,r_2,r_3)$ in the random case is given by:
\begin{eqnarray}
dN = 4\pi r_3^2 dr_3\rho_o \cdot 2\pi r_1\sin(\theta)r_1 d\theta dr_1 \rho_o,
\label{eq:triple1}
\end{eqnarray}
where $\rho_o$ is the average density of the particles.
It follows from the law of cosines for the triangles that for the fixed $r_1$ and $r_3$
we have: $\sin(\theta) d\theta = (r_2 dr_2)/(r_1 r_3)$.
Therefore (\ref{eq:triple1}) can be rewritten as:
\begin{eqnarray}
dN = 8\pi^2 \rho_o^2 r_1 r_2 r_3 dr_1 dr_2 dr_3\;.
\label{eq:triple2}
\end{eqnarray}
Let us assume that $\rho_o = 1/a^3$, where $a$ is the average separation between the particles.
Then we define:
\begin{eqnarray}
\tilde{r}\equiv r/a\;\;\;\;\;\;\textrm{and}\;\;\;\;\;\; g_2(\tilde{r})\equiv \rho(r)/\rho_o.
\label{eq:grdef}
\end{eqnarray}
With these definitions we rewrite (\ref{eq:triple2}) as:
\begin{eqnarray}
dN = 8\pi^2 \tilde{r}_1 \tilde{r}_2 \tilde{r}_3 d\tilde{r}_1 d\tilde{r}_2 d\tilde{r}_3.
\label{eq:triple2scaled}
\end{eqnarray}

Let us consider now a non-random distribution of particles characterized by the pair density function $\rho(r)$.
One simple way to generalize expression (\ref{eq:triple2}) for this case is to assume that:
\begin{eqnarray}
dN = 8\pi^2 \left(\rho_o^2 r_1 r_2 r_3 dr_1 dr_2 dr_3\right) \left[\frac{\rho(r_1)\rho(r_2)\rho(r_3)}{\rho_o^3}\right].
\label{eq:triple3}
\end{eqnarray}
Expression (\ref{eq:triple3}) essentially represents the Kirkwood's superposition
approximation (KSA) \cite{kirkwood1935statistical,hansen1990theory,grouba2004superposition}.
Using the reduced parameters (\ref{eq:grdef}), we rewrite (\ref{eq:triple3}) as:
\begin{eqnarray}
dN = 8\pi^2 \left( \tilde{r}_1 \tilde{r}_2 \tilde{r}_3 d\tilde{r}_1 d\tilde{r}_2 d\tilde{r}_3\right) \left[g_2(\tilde{r}_1)g_2(\tilde{r}_2)g_2(\tilde{r}_3)\right].
\label{eq:triple4}
\end{eqnarray}

It follows from (\ref{eq:triple4}) that it is reasonable to define the triple correlation function (TCF) in the following way:
\begin{eqnarray}
g_3(\tilde{r}_1, \tilde{r}_2, \tilde{r}_2) \equiv \frac{dN}{8\pi^2 \tilde{r}_1 \tilde{r}_2 \tilde{r}_3 d\tilde{r}_1 d\tilde{r}_2 d\tilde{r}_3},
\label{eq:tcfdef}
\end{eqnarray}
where $dN$ is the number of triangles, involving a chosen particle, with the side lengths
in the intervals $(\tilde{r}_1,\tilde{r}_1 + d\tilde{r}_1)$, $(\tilde{r}_2,\tilde{r}_2+ d\tilde{r}_2)$,
and $(\tilde{r}_3,\tilde{r}_3+d\tilde{r}_3)$.

With definition (\ref{eq:tcfdef}), within the KSA, we should have, according to (\ref{eq:triple4}):
\begin{eqnarray}
g_3(\tilde{r}_1,\tilde{r}_2,\tilde{r}_3) = g_2(\tilde{r}_1)g_2(\tilde{r}_2)g_2(\tilde{r}_3).
\label{eq:ksa1}
\end{eqnarray}

In our considerations of the PDF and TCF, we will use reduced units for the distances, $\tilde{r}$, 
to address the structural changes beyond the changes associated with the change in density.
Therefore, in the following, for the briefness of the notations, we will omit upper tildes in the arguments
of these functions.

Often, the TCF defined as the ratio of the TCF in Eq.~\ref{eq:tcfdef} 
to the KSA TCF in Eq.~\ref{eq:ksa1} \cite{raveche1972three,bhatia1976triplet,haymet1981triplet,
haymet1985orientational,stillinger1988theoretical,muller1993triplet,ZahnK2003,Ru2003,grouba2004superposition}. 
Thus defined TCF allows addressing the quality of the KSA.
At present, it is well-known that it does not perform well. 
The purpose of our work is somewhat different, i.e., to address the structural similarity between the
parent-liquid and children-crystalline states. 
Thus, for our purposes, the simpler definition (\ref{eq:tcfdef}) seems to us more suitable.

\subsection{Supplementary Materials II. The triple correlation functions.
\label{sec:suppl-1-triple}\\}

In this Supplementary Materials II (SM2), for illustrative purposes, 
we describe the general shape of the triple correlation function 
for the two simple archetypical systems, i.e., for the random system 
of particles and the random system of particles with an exclusion distance.

\subsubsection{Introduction for the Supplementary materials II.}\label{s:introsupp2}

In Fig.~3 of the paper, we present the results for the triple correlation functions (TCFs) of the studied system.
This figure, in particular, shows the dependencies of TCFs on the lengths of two sides of the triangles, while one side of the triangles has the fixed lengths.

The triangle's inequality largely determines the general shape of the shown TCFs.  This evident inequality states that, for any triangle, the length of any one side must be smaller (or equal) than the sum of the lengths of two other sides.

Although the shapes of the TCFs shown in Fig.~3 could be quite easily understood, we think it is reasonable to discuss them in some detail to clarify the issue and address one small caveat associated with the chosen representation of the TCF.

For these purposes, we consider two archetypical model systems, i.e.,
the system with the completely random distribution of particles and 
the random model with an exclusion distance.

\subsubsection{The TCF for the system with the completely random distribution of point-like particles}
\begin{figure}
\begin{center}
\includegraphics[angle=0,width=3.0in]{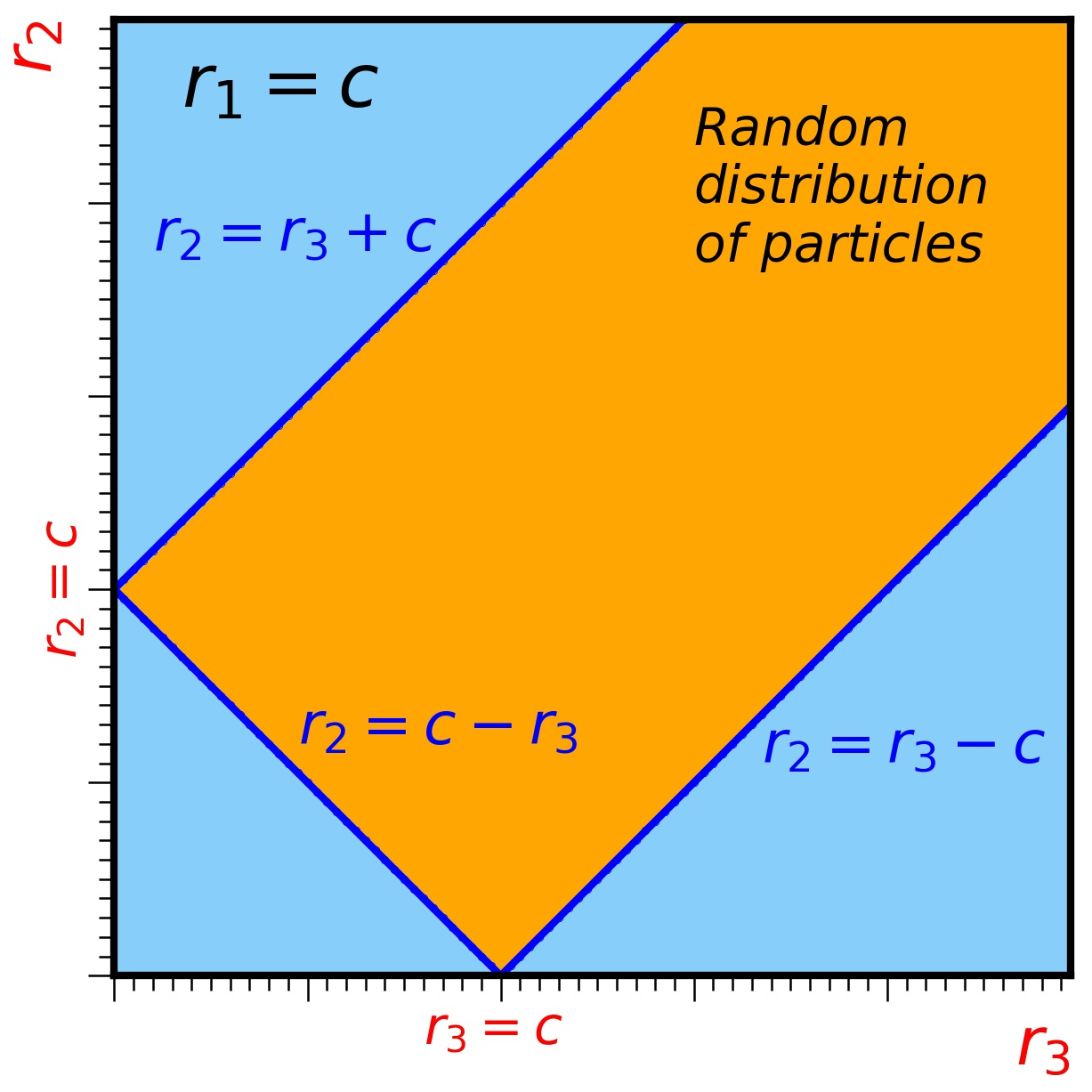}
\caption{\\
The TCF for the completely random distribution of particles. 
The length of one side of the considered triangles is constant: $r_1 = c$.
The orange region corresponds to those triangles that can exist, i.e., to those triangles whose geometries satisfy the triangle's inequality.\\
}\label{fig:TCFs1}
\end{center}
\end{figure}

Figure~\ref{fig:TCFs1} shows the TCF for the case of the random distribution of point-like particles.
The length of one side of the triangles considered in Fig.~\ref{fig:TCFs1} is fixed: $r_1 = c$.

It is easy to realize that the TCF can be non-zero only within the orange area of the figure because there the triangle's inequality is satisfied. The triangles with the geometries outside of the orange region can not exist.

The borders of the orange region correspond to the degenerate triangles for which the sum of the two smaller sides is equal to the length of the largest side. 
For the fixed $r_1 = c$, the equations of these lines are 
shown in the figure.

Note, in Fig.~\ref{fig:TCFs1}, the obvious: when $r_2 = 0$ then $r_3 = c$ and when $r_3 = 0$ then $r_2 = c$.

\subsubsection{The TCF for the random system with an exclusion distance}

\begin{figure}
\begin{center}
\includegraphics[angle=0,width=6.0in]{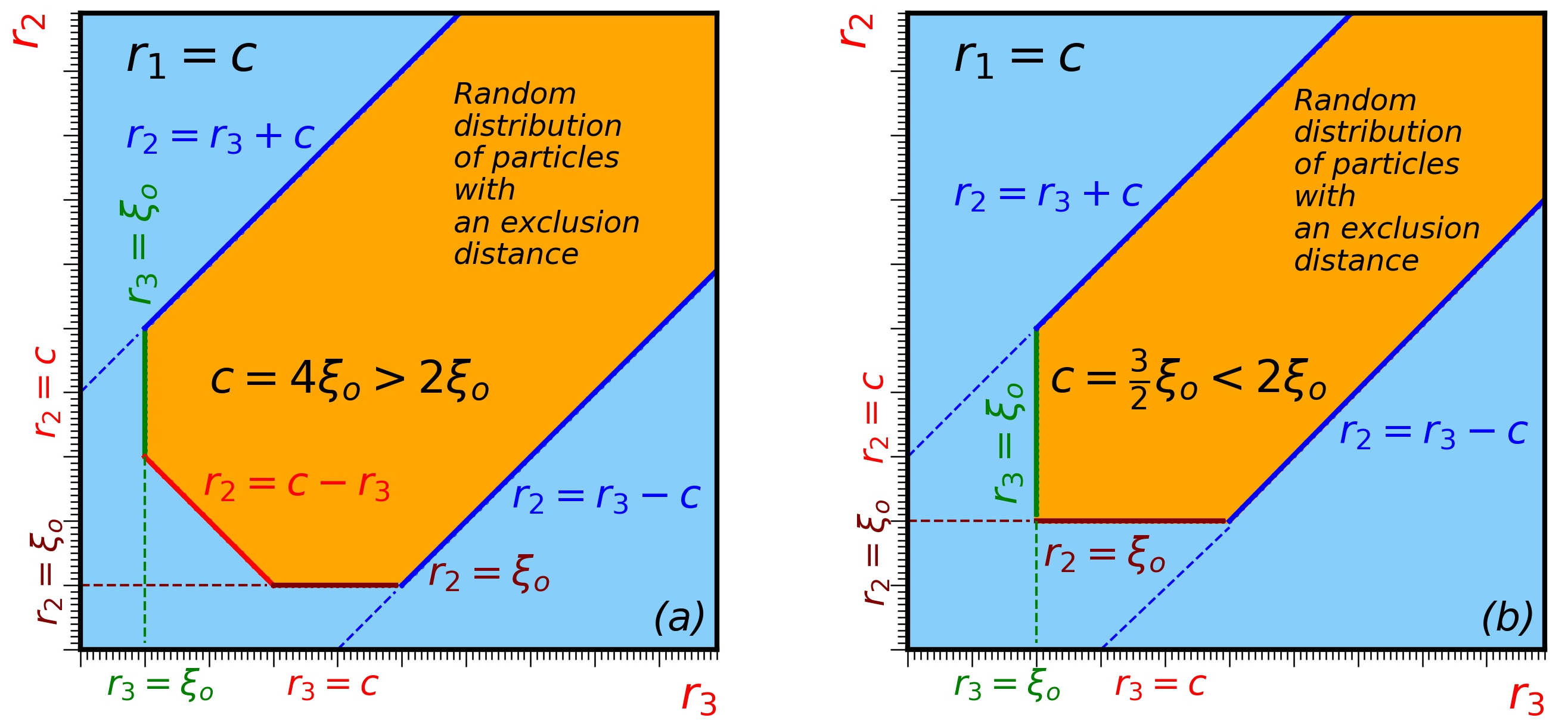}
\caption{\\
In both panels, the orange regions include those parameters of the triangles that satisfy the triangle's inequality,  i.e.,  the orange regions correspond to those triangles that can exist. 
Thus, the TCFs can be non-zero only within the orange regions.\\
Panel (a) corresponds to the case when $c > 2\xi_o$.\\
Panel (b) corresponds to the case when $c < 2\xi_o$.\\
}\label{fig:TCFs2}
\end{center}
\end{figure}

Figure~\ref{fig:TCFs2} shows the TCFs for the two cases of the random distribution of particles with an exclusion distance: $\xi_o$.
The length of one side of the triangles considered in Fig.~\ref{fig:TCFs2} is fixed: $r_1 = c$.

In Fig.~\ref{fig:TCFs2}, the TCFs can be non-zero only within the orange regions. 
Outside of these regions, the TCFs are equal to zero.

For the triangles whose all sides are larger than the exclusion distance, $\xi_o$, the existence of this exclusion distance is irrelevant. 
For these triangles, the equation $r_2 = r_3 - c$ describes the lower bound associated with the corresponding degenerate triangles. 
The lower-diagonal blue lines in both panels of Fig.~\ref{fig:TCFs2} show this bound.
Similarly, the upper bound, associated with the triangle's inequality is given by the equation $r_2 = r_3 + c$ in both panels.

The situation is different for the triangles whose one side is equal to the exclusion distance.
These are the triangles, in Fig.~\ref{fig:TCFs2}(a,b), whose two sides have fixed lengths, i.e., $r_1 = c$ and, for example, $r_2 = \xi_o$. 
These triangles are associated with the maroon lines in Fig.~\ref{fig:TCFs2}(a,b).
The right edges of these lines are associated with the degenerate triangles for which we have:
$r_3 = r_1 + r_2 = c + \xi_o$.

The caveat that we mentioned in the introductory part of this SM is related to the lower allowable bound for $r_3$.
There are two cases:\\

In the 1st case $r_1 = c > 2\xi_o$.\\ 
This case is presented in Fig.\ref{fig:TCFs2}(a).
In this case, both $r_2$ and $r_3$ can not be equal to $\xi_o$ because of the triangles inequality $r_2 + r_3 \geq r_1$.

In the 2nd case $r_1 = c < 2\xi_o$.\\
This case is presented in Fig.\ref{fig:TCFs2}(b).
In this case, both $r_2$ and $r_3$ can be equal to $\xi_o$ because the triangles inequality $r_2 + r_3 \geq r_1$ can be satisfied.

\subsubsection{Conclusion for the Supplementary Materials II}

In our paper, we present the results for the TCF of a soft matter system. 
In this case, the behavior of the TCF is, of course, more complex than 
in the simple examples discussed above. However, because of the generality 
of the triangle's inequality, the shapes of the TCF that we discussed above 
are also relevant for the Fig.~3 of the article.


\bibliographystyle{model1-num-names}


\bibliography{levashov-SI-rsc-arxiv}


\end{document}